\documentclass{amsart}

\usepackage{cite}
\usepackage[utf8]{inputenc}
\usepackage{lineno}
\usepackage[ruled,vlined]{algorithm2e}
\usepackage[dvipsnames]{xcolor}
\usepackage{amssymb}
\usepackage{subfig}
\usepackage{graphicx}
\usepackage{caption} 
\usepackage{textgreek}
\usepackage{amsmath}
\usepackage{amssymb}
\usepackage{bbm}

\DeclareMathOperator{\EX}{\mathbb{E}}
\DeclareMathOperator{\RX}{\mathbb{R}}
\DeclareMathOperator{\ONEX}{\mathbbm{1}} 

\theoremstyle{definition}

\theoremstyle{remark}

\numberwithin{equation}{section}

\begin{document}

\title{Denoising Click-evoked Otoacoustic Emission Signals by Optimal Shrinkage}


\author{Tzu-Chi~Liu}\thanks{T.-C. Liu: Department of Electrical Engineering, National Tsing Hua University, Hsinchu City, 30013, Taiwan. Email: s102061556@m102.nthu.edu.tw}
\author{Yi-Wen Liu}\thanks{Y.-W. Liu: Department of Electrical Engineering, National Tsing Hua University, Hsinchu City, 30013, Taiwan. Email: ywliu@ee.nthu.edu.tw}
\author{Hau-Tieng Wu}\thanks{H.-T. Wu: Department of Mathematics and Department of Statistical Science, Duke University, Durham, NC 27708, United States. Email: hauwu@math.duke.edu}



\date{\today}

\begin{abstract}
Click-evoked otoacoustic emissions (CEOAEs) are clinically used as an objective way to infer whether cochlear functions are normal. However, because the sound pressure level of CEOAEs is typically much lower than the background noise, it usually takes hundreds, if not thousands of repetitions to estimate the signal with sufficient accuracy. In this paper, we propose to improve the signal-to-noise ratio (SNR) of CEOAE signals within limited measurement time by \emph{optimal shrinkage} (OS) in two different settings: the covariance-based OS (cOS) and the singular value decomposition (SVD)-based OS (sOS). By simulation and analyzing human CEOAE data, the cOS consistently reduced the noise and enhanced the SNR by 1 to 2 dB from a baseline method (BM) that is based on calculating the median. The sOS achieved an SNR enhancement of 2 to 3 dB in simulation, and demonstrated capability to enhance the SNR in real recordings when the SNR achieved by the BM was below 0 dB. An appealing property of OS is that it produces an estimate of every individual column of the signal matrix. This property makes it possible to investigate CEOAE dynamics across a longer period of time when the cochlear conditions are not strictly stationary.
\end{abstract}

\maketitle

\section{Introduction}
Otoacoustic emissions (OAEs) refer to sounds coming out from the cochlea. OAEs can be measured in the external ear canal, and have been accepted as a tool for assessing the power-amplifying function in the cochlea associated with the outer hair cells\cite{kemp1986_2}. In clinics, OAEs have been used for detection of hearing loss \cite{hall1999,collet1989}, for examining the hearing of neonates \cite{moleti2005, tognola2005}, and for monitoring the functionality of the cochlea under noise exposure \cite{hotz1993}. For scientific research purposes, a reduction in OAE sound pressure level can be used as an indicator of medial olivocochlear reflex activation \cite{guinan2003, abdala2009}. The generation mechanism of certain types of OAEs is still under debate up to this date (e.g., \cite{Goodman2020}), and models of cochlear mechanics have been built to simulate OAEs \cite{liu2010,Verhulst2012,Bowling2019, liu2016, Vencovsky2020} and explain puzzling phenomena observed in experiments.

The click-evoked otoacoustic emission (CEOAE), a type of OAEs elicited by short acoustic impulses, includes response from different places along the cochlea due to the use of broadband stimuli \cite{marchesi2013}. Because the {\em frequency-to-place mapping} in the cochlea 
is organized in such a way that the characteristic frequency decreases from the base to the apex, CEOAE has a chirp-like waveform in which the high frequency components occur earlier than low frequency components. Thus, CEOAE can be regarded as if it \emph{gathers information} from all along the cochlea. However, to our knowledge, its clinical usage is mainly confined to fast screening of hearing loss due to two problems: (i) it is hard to analyze CEOAE even the signal is clean because of the fast amplitude and frequency modulation and multiple reflections in the cochlea \cite{talmadge1998}, and (ii) the signal-to-noise ratio (SNR) is usually low due to limited measurement time. 

The first challenge can be tackled by modern time-frequency (T-F) analysis techniques \cite{biswal2018}. For instance, the continuous wavelet transform (CWT) was applied 
for visualizing the composition and frequency variation of CEOAEs \cite{notaro2007,tognola1997}, and a modern T-F analysis tool called concentration of frequency and time (ConceFT) can enhance the clarity of CEOAE traces on the T-F plane \cite{wu2018}. 

\begin{figure}
\centering
\includegraphics[width=10cm]{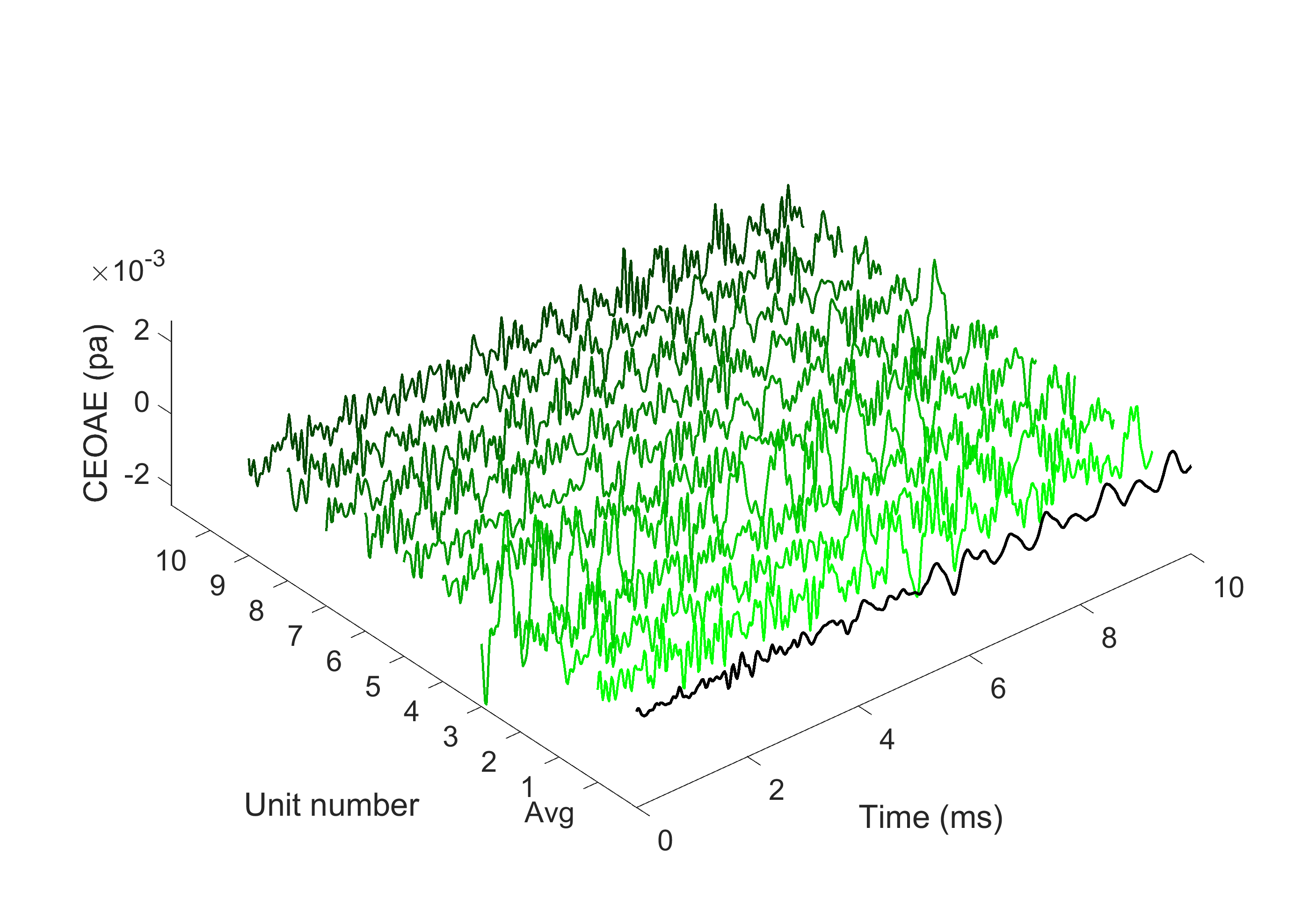}
\caption{\label{fig:10units} An example of CEOAE signal. The first row (Avg) is the median over 1943 recorded units, and the following 10 rows are signals of individual units.}
\end{figure}

The second problem is not much of an issue if the signal quality can be maintained by measuring CEOAEs in a well controlled environment --- the SNR can be enhanced simply by lengthening the recording time. A typical recording protocol recommends that 260 repeating units can be used, each containing four clicks, and the click interval is 20 ms \cite{notaro2007,sisto2007,moleti2005,jedrzejczak2012,ravazzani1998}. Thus, the recording time would be 20.8 s in total. We collected CEOAE signals in normal hearing ears (see Sec.~\ref{sec:bigSec2} for details), and Fig.~\ref{fig:10units} shows an example. Though each individual unit is noisy, the desired CEOAE could be obtained by taking the median over nearly 2000 units. However, for signal acquisition outside a sound booth, the acoustic environment may not be well controlled. Therefore, noise reduction becomes a key step toward successful deployment of CEOAE measurement to less ideal acoustic environments, such as at home, for tele-health purposes.

If the desired signal is known to be narrow-band, a band-pass filter can be utilized to keep components within its frequency range and reject the noise outside. However, the frequency range for CEOAEs is wide, spanning across a few octaves, and excluding all the noise by Fourier-transform-based filtering is impossible. Alternatively, since usually multiple copies of CEOAE signals are collected, it is natural to speculate that by ``integrating'' these signals, the desired signal can be identified. For example, it is a common practice to take the mean or median of all collected CEOAE signals to obtain a high SNR signal \cite{kemp1978}. Further than that, Ravazzani et al. applied principal component analysis (PCA) on the data matrix consisting of all CEOAEs recorded at different stimulus levels 
\cite{ravazzani2003}; 
briefly speaking, the average of repetitions was calculated first for each level, and then PCA was applied on the matrix containing these averages. The first two principal components were kept, and results indicated that the required repetitions of clicks could be reduced to 60 units measured at 3 different stimulus levels. 

With hindsight, the above-mentioned results are reasonable because CEOAEs recorded from the same ear show similar features at different stimulus levels \cite{ravazzani1993,shera2001,sisto2019} that can be effectively captured by PCA. However, due to the well known {\em large p large n}, or high dimensional noise effect in PCA \cite{johnstone2006high}, the recovery of the desired CEOAE hidden in each recording unit might be distorted. We may need to handle this issue when we extract the desired CEOAE from a noisy signal by reducing the noise in the signal matrix.

In this work, we attempt to denoise CEOAE signals by \emph{optimal shrinkage} (OS), and the performance is compared against Wiener filtering (WF) \cite{wiener1950}. OS is a method to recover low-rank matrices from noisy data by shrinkage of the eigenvalues of the covariance structure \cite{donoho2018optimal} or the singular values of the data matrix \cite{gavish2017} based on the chosen optimization criterion. The framework of OS works if the rank of signal matrix is low. For CEOAEs, since CEOAEs recorded from the same ear are ``similar'', it is reasonable to assume that the number of components, which corresponds to the rank of its signal matrix, is limited. Therefore, with the assumption that the noise statistics stays invariant, both OS and WF are suitable for denoising CEOAE signals. The difference between these two approaches will be investigated empirically in this work.

The paper is organized as follows. In Sec.~\ref{sec:bigSec2}, methods for CEOAE data collection are described and the mathematical details of OS and WF are introduced. The comparison of denoising methods for both simulated and measured CEOAEs is shown in Sec.~\ref{sec:bigSec3}. Discussions and conclusions are given in Sec.~\ref{sec:bigSec4} and Sec.~\ref{sec:bigSec5}, respectively.

\section{Materials and methods} \label{sec:bigSec2}
In this section, details of the data collection are introduced first, including the recording equipment, the subjects, and design of the stimulus. Then we describe the signal pre-processing method, including signal segmentation and artifact rejection. Afterwards, the implementation for two matrix denoising methods, WF and OS, is presented. The last part describes the way to generate CEOAE signals by simulation of cochlear mechanics for evaluating the denoising methods. 

\subsection{CEOAE measurement}
CEOAEs were recorded by an ER-10X microphone (Etymotic Research Inc., Elkgrove Village, IL, USA) and a RME Fireface UFX II soundcard (RME, Haimhausen, Germany) in a sound-proof room at National Tsing Hua University, Taiwan. The method for measurement is referred to as a ``nonlinear protocol'' \cite{kemp1986} by which the initial linear response not pertaining to cochlear conditions could be eliminated \cite{jedrzejczak2012}. In this method, a unit of the stimulus contains three clicks, in which the first two clicks are equal in amplitude and the third has the opposite sign and its amplitude is two times larger. Then the CEOAE is extracted, according to its nonlinear growth with the stimulus level, by summing the responses to the three clicks \cite{withnell2005}. 

The interval between two clicks was about 99.8 ms (4400 sample points in 44.1 kHz sampling rate), the peak sound pressure level of the first two clicks in a unit was set to 76 dB sound pressure level (SPL), and 2000 units were presented for each recording session. The click interval was set sufficiently long so that synchronized spontaneous otoacoustic emissions (SSOAEs) could also be observed in the recordings \cite{prieve1995,jedrzejczak2008}. Although the click interval was about 100 ms, only the response during the first 20 ms after the click was analyzed in this study.

Eight subjects of age 21-25 years were recruited. Overall, 14 ears were successfully measured three times across different days; data from 2 ears were excluded due to artifact rejection (details are described in the next section). Thus, 42 recordings were obtained in total, 
and they are referred to as file\#1 to file\#42 in this paper. 

\subsection{Segmentation and signal quality determination}
A band-pass filter was applied on the recorded signal to reject frequencies below 0.8 kHz and above 10 kHz. Subsequently, the first local maximum was regarded as the first click, and it was considered as time zero. Then the signal was divided into non-overlapping units of length $3T$, where $T\approx$ 99.8 ms denotes the interval between clicks. Thus, each unit contained two positive clicks and one negative click of twice the amplitude. For CEOAE extraction under the nonlinear protocol, responses to the three clicks were summed up as previously described.

Besides environmental noise, human generated noise such as swallowing and friction sounds could also be found in the signal. To maintain the signal quality for analysis, a unit would be rejected if any instantaneous amplitude after 5 ms from time zero was higher than 50 dB SPL. If more than 80\% units were rejected, the file would be abandoned; if one file out of three measurements was abandoned, the corresponding ear would be neglected, and that was why only 14 ears out of 16 were analyzed. In average, 1940 units remained after artifact rejection.

\subsection{Noise reduction algorithms} \label{sec:noise_reduction}
After segmentation and artifact rejection, the responses are denoted as $y_i\in \RX^p$, where $i=1,\ldots,n$, $p=882$ (corresponding to a recording over 20 ms when the sampling rate is 44.1 kHz) is the length of the signals, and $n$ denotes the number of units. Thus a $p$ by $n$ matrix $Y$ is formed, where the $i$th column is $y_i$. Then, a matrix denoising method such as WF or OS could be applied to estimate the desired signal matrix $X$. In particular, we are interested in knowing how the denoising performance varies with $n$ when $p$ is fixed.


\subsubsection{Noise reduction by cOS} \label{sec:covOS}
We assume that $y_i$ satisfies
\begin{equation} \label{eq:YformC}
    y_i=x_i+\sigma \xi_i,
\end{equation}
where $x_i \in \RX^p$ is the desired {\em clean} CEOAE signal satisfying $\EX x_i=\mu$ and $\EX (x_i-\mu)(x_i-\mu)^\top=\Sigma$, $\sigma\geq 0$ is the standard deviation of noise and the noise $\xi_i\in \RX^p$ satisfies $\EX \xi_i \xi_i^\top=I_{p\times p}$ with finite fourth moment. We assume that $x_i$ and $\xi_i$ are independent, and $\{y_i\}$ are i.i.d. generated from \eqref{eq:YformC}. 

We follow the procedures described in the literature \cite{singer2013} to apply the cOS on the matrix $Y$. Note that this is a special case of those considered in \cite{donoho2018optimal} when the Frobenius norm is considered in the optimization step for the purpose of recovering the covariance matrix. It is possible to consider other norms, like operator norm or nuclear norm in the optimization step. We refer readers with interest to \cite{donoho2018optimal} for details. First, the mean of the desired signal matrix is estimated as
\begin{equation}
    \hat{\mu}=\frac{1}{n}\sum_{i=1}^n y_i.
\end{equation}
Afterwards, the empirical covariance matrix of the observed signal is calculated as follows,
\begin{equation} \label{eq:Sn}
    S_n=\frac{1}{n}\sum_{i=1}^n (y_i-\hat{\mu})(y_i-\hat{\mu})^\top.
\end{equation}
The eigenvalues of $S_n$ are denoted as $l_1\geq l_2 \geq ... \geq l_p \geq 0$, and the corresponding eigenvectors are denoted as $\{u_1,u_2,...,u_p\}$, respectively. Then the relation between the eigenvalues of the desired signal matrix $X$, denoted as $\{\lambda_1,\lambda_2,...,\lambda_p\}$, and the eigenvalues of $S_n$ is given by the following quadratic equation \cite{singer2013},
\begin{equation} \label{eq:quadratic}
    l_i=(\lambda_i+\sigma^2)\left(1+\beta\frac{\sigma^2}{\lambda_i}\right),
\end{equation}
where $\beta=p/n$ and $\sigma$ is the standard deviation of noise. Note that the length of an unit is about 99.8 ms and the first 20 ms after the click contains the CEOAE signal. For noise estimation, the last 40 ms of the $i$-th unit is assumed to be the noise signal, denoted as $w_i$. Then the estimation of $\sigma$ is achieved by
\begin{equation} \label{eq:sigma}
    \sigma=\frac{1}{n}\sum_{i=1}^n\sigma_{w,i},
\end{equation}
where $\sigma_{w,i}$ is the standard deviation of $w_i$. We mention that while there exists more sophisticated noise estimation algorithm like \cite{kritchman2008determining}, they are limited when applied to real data. We thus focus on this simple noise estimation in this work. Afterwards, the positive root of Eq.~\ref{eq:quadratic} is selected as the estimated $\lambda_i$; that is
\begin{equation}
    \lambda_i=\frac{-[\sigma^2 (1+\beta)-l_i]+\sqrt{[\sigma^2 (1+\beta)-l_i]^2-4\sigma^4\beta}}{2}.
\end{equation}

The filter coefficients can be written as
\begin{equation}
    h_i=\frac{K_{\beta,i}}{1+K_{\beta,i}},
\end{equation}
where $i=1,...,p$ and
\begin{equation} \label{eq:definition_Kbetai}
    K_{\beta,i}=\left\{ \begin{array}{rcl}
    0 & \mathrm{for} & \lambda_i/\sigma^2 < \sqrt{\beta} \\
    \frac{(\lambda_i/\sigma^2)^2-\beta}{\lambda_i/\sigma^2+\beta} & \mathrm{for} & \lambda_i/\sigma^2 \geq \sqrt{\beta}
    \end{array} \right.
    .
\end{equation}
Finally, the desired signal matrix is estimated as
\begin{equation} \label{eq:covOSform}
    \hat{X}=\hat{\mu}\ONEX^\top+UHU^\top(Y-\hat{\mu}\ONEX^\top),
\end{equation}
where $U=[u_1,u_2,...,u_p]\in \RX^{p\times p}$, $H$ denotes a diagonal matrix whose $i$th diagonal element is $h_i$, and $\ONEX=[1,1,\ldots,1]^\top\in \RX^p$. We mention that the cOS is a filtering technique that is more aggressive than the WF. Indeed, in the WF, the eigenvalues are not suppressed as aggressively as that shown in $H$. See \cite{singer2013} for a more detailed discussion of this relationship and a summary of the algorithm in Sec.~\ref{sec:WF}.

\subsubsection{Noise reduction by sOS} \label{sec:svdOS}
We assume that $Y$ satisfies
\begin{equation}\label{eq:Yform}
    Y=X+\sigma Z,
\end{equation}
where $X\in\RX^{p\times n}$ consists of the desired {\em clean} CEOAE signal in the columns and we assume that $X$ a low-rank matrix, $\sigma\geq 0$ is the standard deviation of noise and the noise matrix $Z\in \RX^{p\times n}$ has i.i.d. entries with zero mean, unit variance and finite fourth moment. Compared with \eqref{eq:YformC}, where the variation of CEOAE signals is quantified by the covariance structure, in \eqref{eq:Yform}, the variation of CEOAE signals is quantified in the mean. The goal is to estimate the matrix $\hat{X}$, or denoise the noisy matrix, via SVD such that the Frobenius norm loss $L^{fro}=\| X-\hat{X}\|_{F}$ is minimized. We mention that SVD is commonly applied in spatiotemporal analysis \cite{onorati2013}.
While other norms, like operator norm or nuclear norm, could also be considered \cite{gavish2017}, in this work we focus on the Frobenius norm to simplify the discussion.

The standard deviation of noise is first estimated as \eqref{eq:sigma}. Then, the singular value decomposition is performed on $\hat{Y}=Y/\sigma\sqrt{n}$ such that
\begin{equation} \label{SVDonY}
    \hat{Y}=ULV^\top,
\end{equation}
where $U$ is a $p\times p$ orthogonal matrix, $V$ is a $n\times n$ orthogonal matrix, and $L\in \RX^{p\times n}$ with the ($i,i$)th entry being the $i$th singular value $\sigma_i$ in the descending order and all other entries being zero.
After that, an optimal shrinker $\eta^*$ is set as \cite{gavish2017}
\begin{equation} \label{eq:shrinker}
    \eta^*(\xi)=\left\{
    \begin{array}{ll}
        \frac{1}{\xi}\sqrt{(\xi^2-\beta-1)^2-4\beta}, &  \xi\geq 1+\sqrt{\beta}\\
        0, & \xi< 1+\sqrt{\beta}
    \end{array}
    \right.
    .
\end{equation}
Note that $\eta^*(\xi)/\xi$ approaches $1$ when $\xi\to \infty$. Finally, the estimation of desired signal matrix is obtained by 
\begin{equation}
    \hat{X}=\sigma\sqrt{n}\cdot U\hat{L}V^\top,
\end{equation}
where $\hat{L}\in \RX^{p\times n}$ is the result of modifying the ($i,i$)th entry of $L$ from $\sigma_i$ to $\eta^*(\sigma_i)$. 

\subsubsection{Noise reduction by WF} \label{sec:WF}
The estimator of the WF has the same form as \eqref{eq:covOSform}, but the eigenvalues are not ``shrinked''. The filter coefficients are defined as
\begin{equation}
    \hat{h}_i=\frac{l_i}{l_i+\sigma^2},
\end{equation}
where $l_i$ is the $i$th eigenvalues of $S_n$ as derived in \eqref{eq:Sn}, and $\sigma$ is estimated as \eqref{eq:sigma}. Then a $p\times p$ diagonal matrix $\hat{H}$ is obtained and its $i$th diagonal element is $\hat{h}_i$. Afterwards, the desired signal is estimated as \eqref{eq:covOSform} with $H$ replacing by $\hat{H}$.

\subsection{Generation of simulated CEOAEs}
Following Wu and Liu \cite{wu2018}, an electromechanical model of cochlea \cite{liu2010} was adopted to generate CEOAE by simulation. The model was based on a transmission-line model of cochlear mechanics \cite{neely1985,neely1993}, and the outer hair cell (OHC) was characterized by a piezo-electrical equivalent circuit \cite{liu2009,mountain1994}. As in \cite{wu2018}, a term called ``roughness'' was added to the model by applying perturbation to the mass density along the basilar membrane, and the level of roughness was adjusted so as to cause coherent reflection of the traveling waves at an empirically reasonable level. The roughness influenced both the shape and the level of CEOAE. For adjusting only the level, the membrane conductance of outer hair cells in the model could be multiplied uniformly by a factor\cite{MoH2014}. A middle-ear model was also attached to the cochlear model, so that the OAE signal could be ``recorded'' in the ear canal. 

\section{Results} \label{sec:bigSec3}
In this section, the behavior of singular values before and after the sOS is first shown. Afterwards, WF and both cOS and sOS are applied to simulated and real CEOAE signals. The results are quantified by SNRs. Instead of the whole estimated matrix $\hat{X}$, its point-wise median along the second axis, denoted as $\hat{x}\in \RX^p$ and its $i$th element being the median of $\{\hat{X}(i,1),...,\hat{X}(i,n)\}$, is visualized and compared for different $n=400,200,100$ and $50$. In our simulations, the ground truth $x$ is the point-wise median of the desired signal matrix $X$ (Sec.~\ref{sec:result_simulation}); otherwise, it would be determined from the recording (Sec.~\ref{sec:result_real}). Either way, the error of the estimated CEOAE $\hat{x}$ is defined as $w=\hat{x}-x$. Afterwards, the SNR is defined as
\begin{equation}
    \mathrm{SNR}=10\mathrm{log}_{10}\frac{\sigma^2_x}{\sigma^2_w},
\end{equation}
where $\sigma_x$ and $\sigma_w$ are standard deviations of $x$ and $w$, respectively.

\subsection{Singular value behavior}
\begin{figure}[ht]
\centering
\includegraphics[width=8cm]{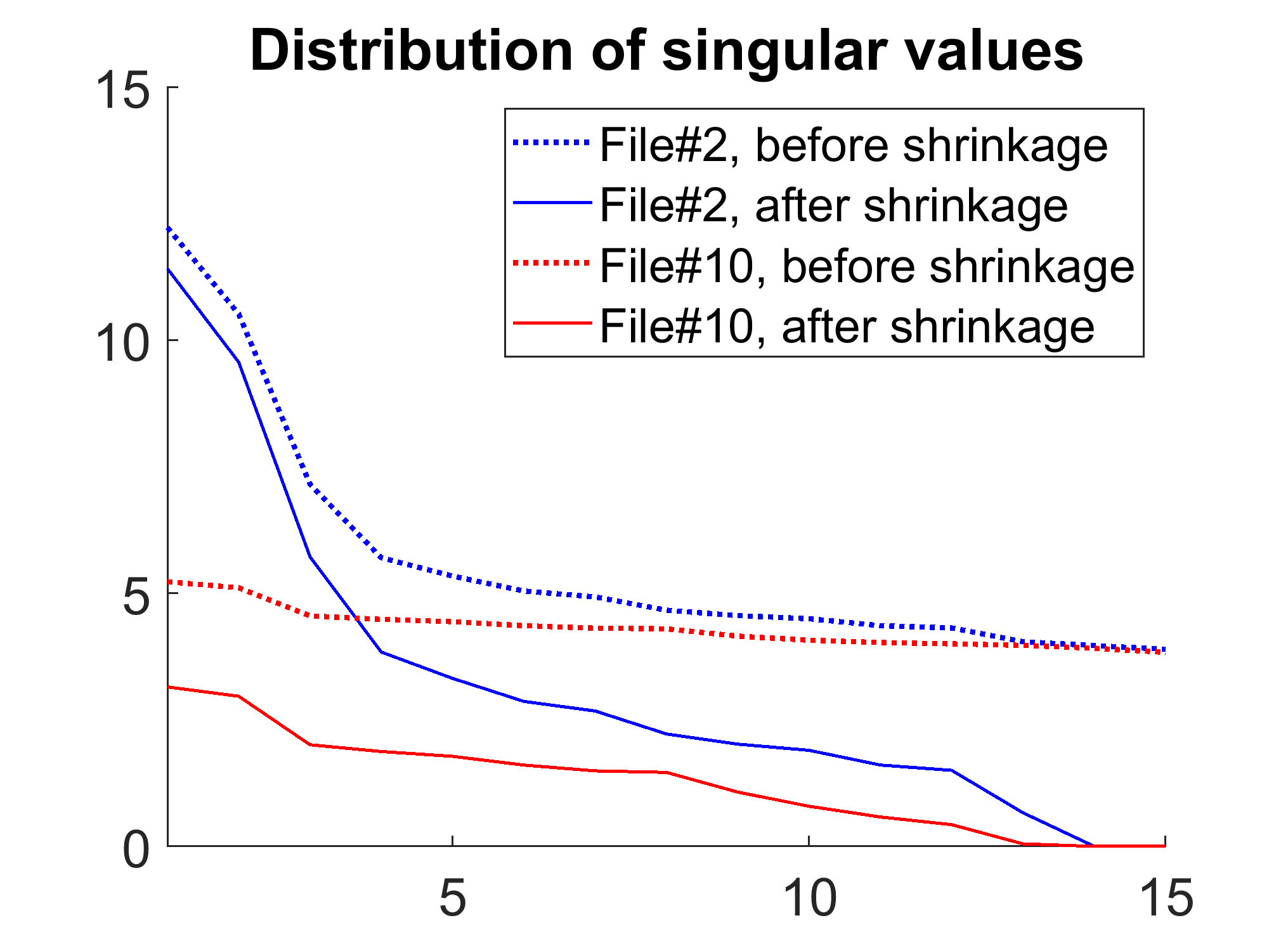}
\caption{\label{fig:singular_real} Singular values before and after OS of two examples: file\#2 and file\#10.}
\end{figure}

A typical example of the distribution of singular values before and after OS is shown in blue line in Fig.~\ref{fig:singular_real}. In this example, the first singular value is sufficiently higher than $1+\sqrt{\beta}$ and it is not reduced much by the optimal shrinker. Due to the nature of OS in \eqref{eq:shrinker}, a smaller singular value is reduced even more than a larger one, and the values become all $0$ after the $14$th of them. The CEOAE signal with $n=100$ is used for this plot, and the details are described in Sec.~\ref{sec:result_real}. 

\subsection{Denoising simulated CEOAE} \label{sec:result_simulation}
In this section, the matrix denoising methods are evaluated on simulated CEOAE data. The generation of the CEOAE signal (length $p$) is repeated $n$ times, and these signals are saved in a $p\times n$ matrix $X$. Here, $p=1280$ (corresponding to 20 ms long signal sampled at 64 kHz sampling rate). The membrane conductance of outer hair cells in the model is multiplied by a random factor that follows $\mathcal{N}(1,0.01)$ and the CEOAE root mean square level is $-1.30\pm 0.17$ dB SPL in the simulation. Then a noise matrix denoted as $\sigma Z$, where $\sigma$ is the standard deviation of noise and $Z\sim \mathcal{N}(0,1)$ has i.i.d. entries, 
is added to the signal matrix as in \eqref{eq:Yform}.

\begin{figure}[t]
\centering
\subfloat[]{
  \includegraphics[width=7cm]{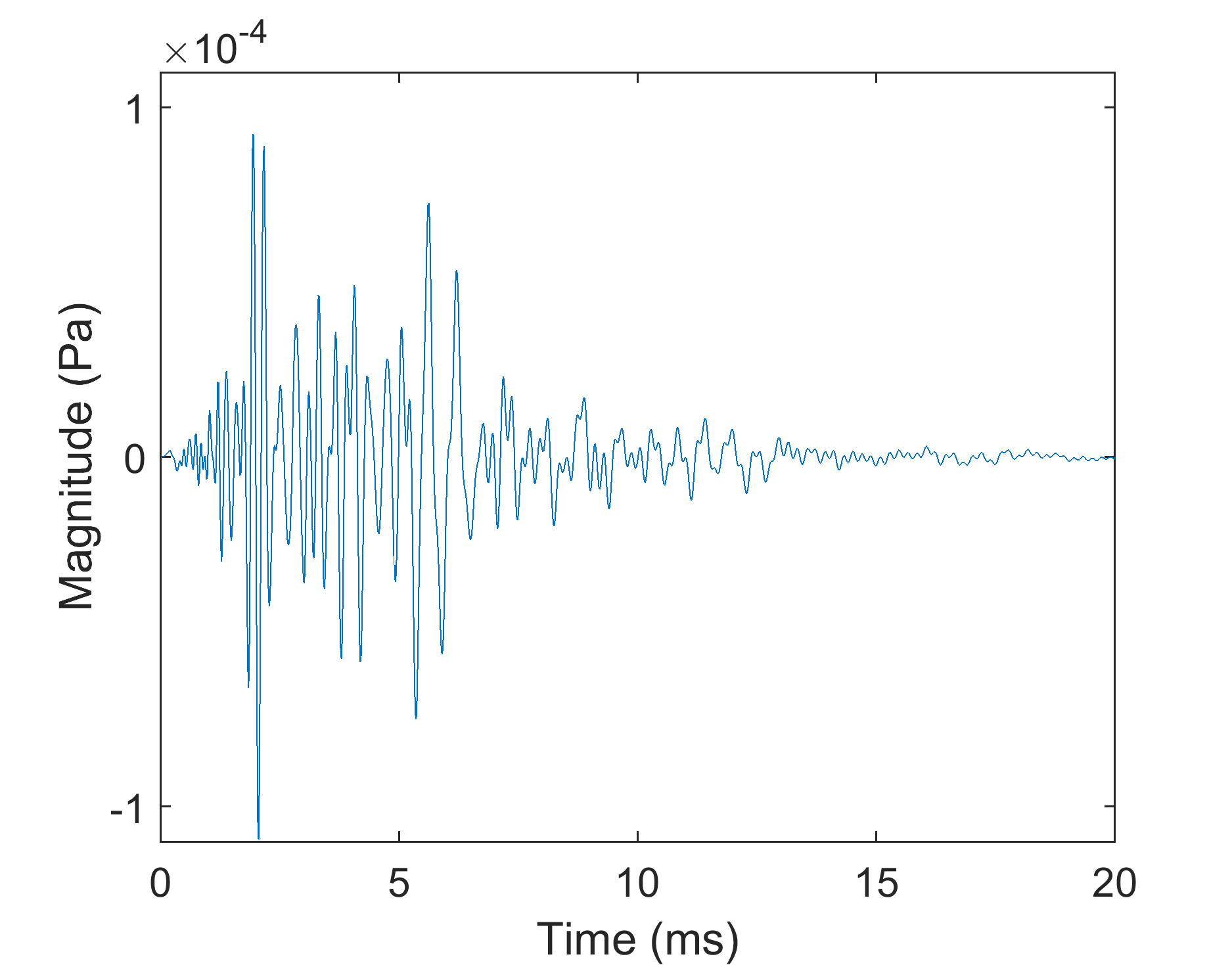}
  \label{fig:conceFTA}
}\quad
\subfloat[]{
  \includegraphics[width=7cm]{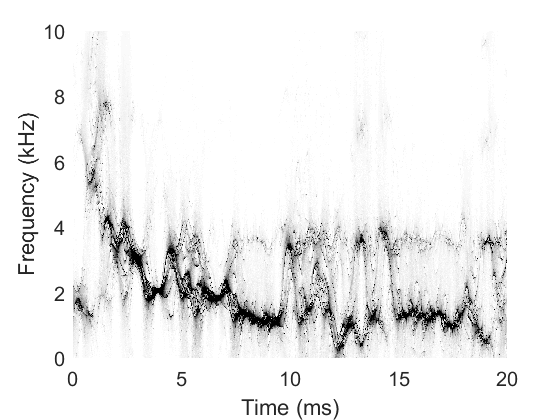}
  \label{fig:conceFTB}
}
\caption{\label{fig:conceFT} The CEOAE generated by the cochlear model. (a) The waveform. (b) The ConceFT T-F representation. The trace on (b) shows that the characteristics of a CEOAE signal are simulated in that the instantaneous frequency decreases from high (about 6 Hz) to low (about 1 kHz) within about 10 ms.}
\end{figure}

A typical waveform of the simulated CEOAE signal is shown in Fig.~\ref{fig:conceFTA}. Note that both the amplitude and the frequency of the signal are fluctuating within the first 20 ms. To extract the fluctuating amplitude and instantaneous frequency, a nonlinear T-F analysis method called ConceFT \cite{daubechies2016} is adopted, and the result is shown in Fig.~\ref{fig:conceFTB}.

Before the comparison between different denoising methods, we compare the usage of different norms for the sOS (see \cite{gavish2017} for the optimal shrinkers for the operator norm and the nuclear norm). 
Here, $n$ varies from 400 to 50, and the value of noise level $\sigma$ is set at 1.93 $\mu$Pa (the reason would be mentioned later).
The result is shown in Table \ref{tab:compare_norm}. It is clear that overall the performance of nuclear norm is the worst, and the performance of Frobenius norm is better when the sample size is small.
Therefore, we choose the Frobenius norm for the sOS in this paper. For cOS, we also choose the Frobenius norm to simplify the discussion.

\begin{table}[h]
\caption{\label{tab:compare_norm} SNR of estimated CEOAEs by the sOS with different norms. Each value is the median of the results in 100 realizations of noise, and inside the parenthesis is the corresponding interquartile range. The unit is dB.}
\begin{center}
\begin{tabular}{c|c|c|c|c}
    \hline
    \hline
    $n$ & 400 & 200 & 100 & 50\\
    \hline
    Frobenius & 7.41 (0.19) & \textbf{5.12 (0.23)} & \textbf{3.23 (0.18)} & \textbf{1.80 (0.22)} \\
    Operator & \textbf{7.45 (0.22)} & 5.02 (0.21) & 2.82 (0.19) & 1.14 (0.36) \\
    Nuclear & 6.89 (0.30) & 4.57 (0.37) & 2.60 (0.38) & 0.92 (0.46) \\
    \hline
    \hline
\end{tabular}
\end{center}
\end{table}

\begin{table}[h!]
\caption{\label{tab:result_sim} SNR of estimated CEOAEs with different kinds of additive noise. Four cases are compared: the signal denoised by the baseline method (BM) of the median of the observed matrix, by Wiener filtering (WF), by covariance-based OS (cOS) and by SVD-based OS (sOS), respectively. The unit and the meaning of each value is similar to Table~\ref{tab:compare_norm}.}
\begin{center}
\subfloat[\label{tab:sim_WGN} White noise]{
\begin{tabular}{c|c|c|c|c}
    \hline
    \hline
    $n$ & 400 & 200 & 100 & 50\\
    \hline
    BM & 5.08 (0.22) & 2.11 (0.20) & -0.90 (0.28) & -3.85 (0.28) \\ 
    WF & 5.71 (0.24) & 2.51 (0.24) & -0.67 (0.25) & -3.74 (0.28) \\
    cOS & 7.02 (0.21) & 4.03 (0.19) & 0.99 (0.23) & -2.01 (0.27) \\
    sOS & \textbf{7.41 (0.19)} & \textbf{5.12 (0.23)} & \textbf{3.23 (0.18)} & \textbf{1.80 (0.22)} \\
    \hline
    \hline
\end{tabular}
} 
\quad
\subfloat[\label{tab:sim_rain} Rain]{
\begin{tabular}{c|c|c|c|c}
    \hline
    \hline
    $n$ & 400 & 200 & 100 & 50\\
    \hline
    BM & 5.53 (1.21) & 2.69 (1.39) & -0.34 (1.41) & -3.50 (1.71) \\ 
    WF & 5.93 (1.14) & 3.00 (1.45) & -0.16 (1.40) & -3.40 (1.69) \\
    cOS & \textbf{6.27 (1.44)} & 3.65 (1.51) & 0.67 (1.31) & -2.38 (1.50) \\
    sOS & 6.02 (1.58) & \textbf{3.66 (1.53)} & \textbf{1.76 (1.14)} & \textbf{0.57 (0.87)} \\
    \hline
    \hline
\end{tabular}
}
\quad
\subfloat[\label{tab:sim_office} Office]{
\begin{tabular}{c|c|c|c|c}
    \hline
    \hline
    $n$ & 400 & 200 & 100 & 50\\
    \hline
    BM & 6.70 (1.79) & 3.73 (1.99) & 0.54 (1.84) & -2.05 (1.98) \\ 
    WF & 6.89 (1.71) & 3.94 (1.88) & 0.75 (1.94) & -1.85 (1.94) \\
    cOS & \textbf{7.08 (1.98)} & \textbf{4.50 (1.87)} & \textbf{1.48 (2.05)} & -1.15 (1.84) \\
    sOS & 6.50 (2.04) & 3.57 (2.13) & 1.01 (1.60) & \textbf{-0.23 (1.30)}\\
    \hline
    \hline
\end{tabular}
}
\quad
\subfloat[\label{tab:sim_restaurant} Restaurant]{
\begin{tabular}{c|c|c|c|c}
    \hline
    \hline
    $n$ & 400 & 200 & 100 & 50\\
    \hline
    BM & 4.35 (1.35) & 1.74 (1.63) & -0.80 (1.02) & -3.57 (1.45)\\ 
    WF & 4.34 (1.45) & 1.91 (1.68) & -0.65 (1.04) & -3.49 (1.44) \\
    cOS & \textbf{4.36 (1.63)} & \textbf{2.00 (1.61)} & \textbf{-0.47 (1.43)}  & -2.95 (1.69) \\
    sOS & 3.77 (1.44) & 0.74 (1.30) & -0.79 (0.83) & \textbf{-1.69 (1.13)} \\
    \hline
    \hline
\end{tabular}
}
\end{center}
\end{table}

Next, we examine how the denoising algorithms depend on the number of observations. Four cases are compared with 100 different realizations of noise: the signal denoised by the baseline method (BM) of taking the median of the observed matrix $Y$ along the second axis, the signal denoised by WF, by cOS, and by sOS, respectively. The setting of $n$ and $\sigma$ is equal to what is described in the previous paragraph. The value of $\sigma$ is determined such that the SNR is positive at $n>100$ and is negative at $n\leq 100$ for the BM. 
On Table~\ref{tab:sim_WGN}, the SNR achieved by WF is higher than that by the BM but the difference is all less than 1 dB, so we focus on 
cOS and sOS hereafter. As a rule of signal averaging, the SNR of the original signal decreases by about 3 dB when $n$ is reduced by 1/2. Although cOS improves the SNR, the improvement is similar under all settings of $n$ and it still follows the 3-dB rule. In contrast, the ``SNR enhancement'' by the sOS increases as $n$ decreases. Thus, the sOS performs better than the cOS in that it shows a more graceful SNR degradation as $n$ decreases.

To further evaluate how ambient noise impacts the OS, besides white Gaussian noise, different types of ambient noise are added on the signal matrix. By doing so, we aim to evaluate the denoising methods more realistically. Ten-minute recordings of background noise from three environments (``raining''\footnote{Retrieved from https://www.youtube.com/watch?v=q76bMs-NwRk}, ``office''\footnote{Retrieved from https://www.youtube.com/watch?v=8tKpjMh\_OUw}, and ``restaurant''\footnote{Retrieved from https://www.youtube.com/watch?v=xY0GEpbWreY}) were retrieved from YouTube, 
resampled to the same sampling rate as the simulated CEOAE (64 kHz), and normalized to zero mean and unit variance. To mimic sequential recording, a period with length $p\times n$ samples is randomly picked from an ambient sound, and a noise matrix $Z\in \RX^{p\times n}$ is formed. Then, a signal matrix is created by $Y=X+\sigma Z$, where the value of $\sigma$ is equal to the simulation using white noise. 

The results of simulation using ambient noise are listed in Table~\ref{tab:sim_rain}, \ref{tab:sim_office} and \ref{tab:sim_restaurant}. Similar to the result of using white noise, we focus on the discussing cOS and sOS. 
Note that, in the rain, the baseline SNR is similar to the white noise, but the SNR decreases 0.5 to 1 dB for the cOS and decreases 1 to 1.5 dB for the sOS. Thus, the SNR achieved by cOS is higher than by sOS when $n=400$. In an office or a restaurant
, the cOS has the best performance at $n=400,200$ and $100$, and the sOS is better than the cOS at $n=50$.

\begin{figure*}[htb!]
\centering
\includegraphics[width=13cm]{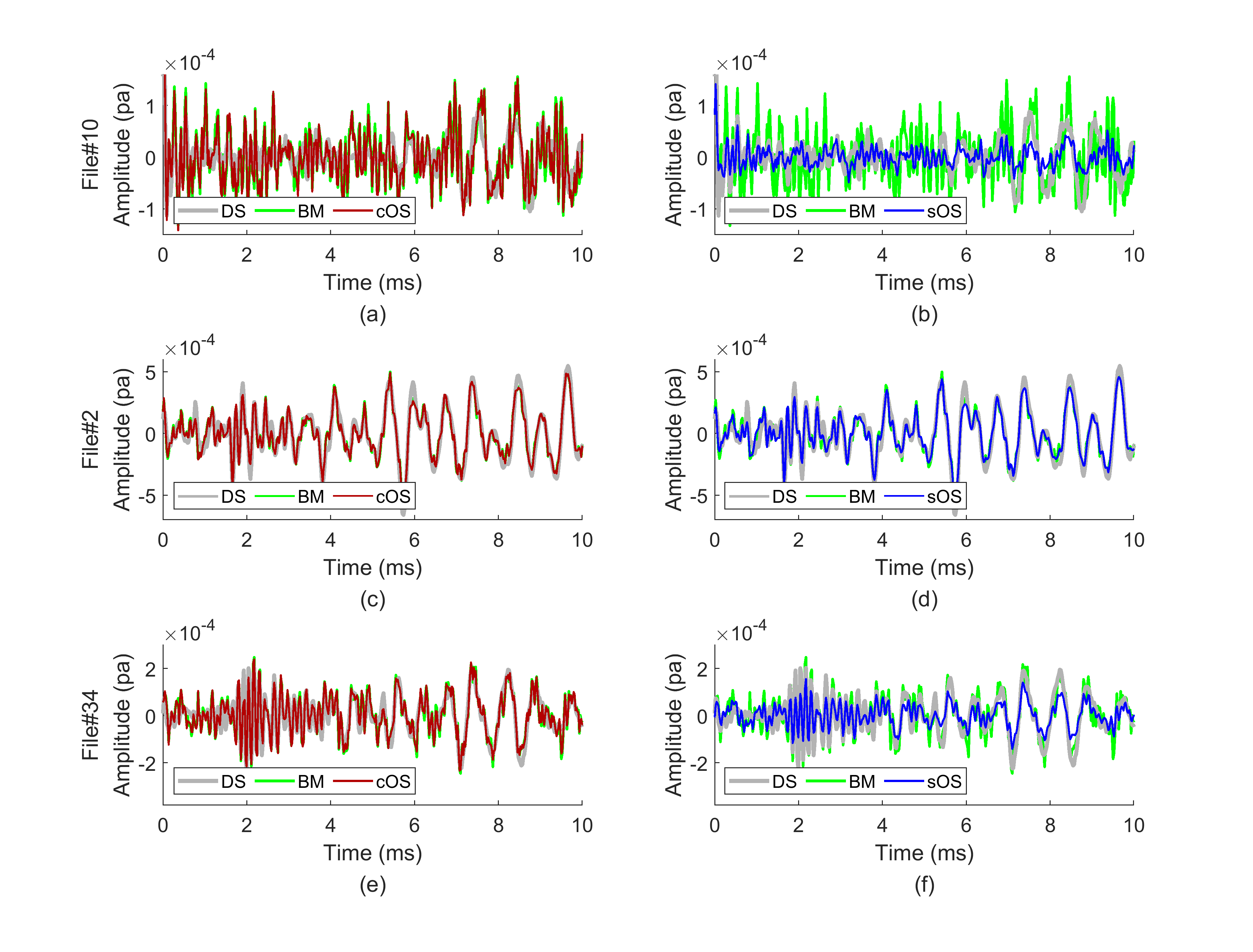}
\caption{\label{fig:ENVreal} Examples of results of denoising when only the first 100 units are analyzed. The desired signal (DS) and the signals denoised by the BM, by cOS and by sOS are compared. (a) and (b): File\#10. (c) and (d): File\#2. (e) and (f): File\#34.
}
\end{figure*}

\subsection{Denoising real CEOAE data} \label{sec:result_real}
In this section, the denoising methods are applied on real CEOAE signals. The number of retained units was $1940.2\pm 40.4$ after artifact rejection among 42 recordings. The clean signal matrix $X$ is not available from real data. We assume that the cochlea can be regarded as a time-invariant system throughout the time course of CEOAE measurement, so the desired result $x$ is defined as the point-wise median of all the retained $y_i$'s in a full recording for the calculation of the SNR. The noise level was $25\pm 0.49$ dB SPL among 42 recordings.

Fig.~\ref{fig:ENVreal} shows some waveform examples. 
Four cases are compared: the desired signal (DS) $x$ estimated from all units, the signal denoised by BM, by cOS, and by sOS, respectively. Note that for the latter three cases, the number of observation $n$ is set at 100, which means only the first 100 units are analyzed. Empirically, we found that WF-based denoising produces a waveform that is quite similar to what is produced by the BM; so the WF results are not shown here.

Fig.~\ref{fig:ENVreal}a and \ref{fig:ENVreal}b shows the denoised result of file\#10. In this case, the SNR is $-2.59$ dB for the BM, and the noise is most successfully reduced by the sOS, achieving 3.95 dB enhancement from the baseline. Before $t=7$ ms, the denoised result by the sOS is similar to the desired signal which is defined as the median of all ($n=1967$) units. Hence for this example, the sOS is able to partially recover the desired CEOAE signal when only about 5\% of the units are available for analysis.

OS does not perform similarly well for every signal. The signals denoised by different methods are similar for file\#2, as shown in Fig.~\ref{fig:ENVreal}c and Fig.~\ref{fig:ENVreal}d. The reason might be that the SNR achieved by BM is already high (10.2 dB) so the matrix denoising methods are of little use to enhance the signal further. Fig.~\ref{fig:ENVreal}e and Fig.~\ref{fig:ENVreal}f will be discussed in Sec.~\ref{sec:special_cases}.

\begin{figure}[ht]
\centering
\subfloat[SNR]{
  \includegraphics[width=9cm]{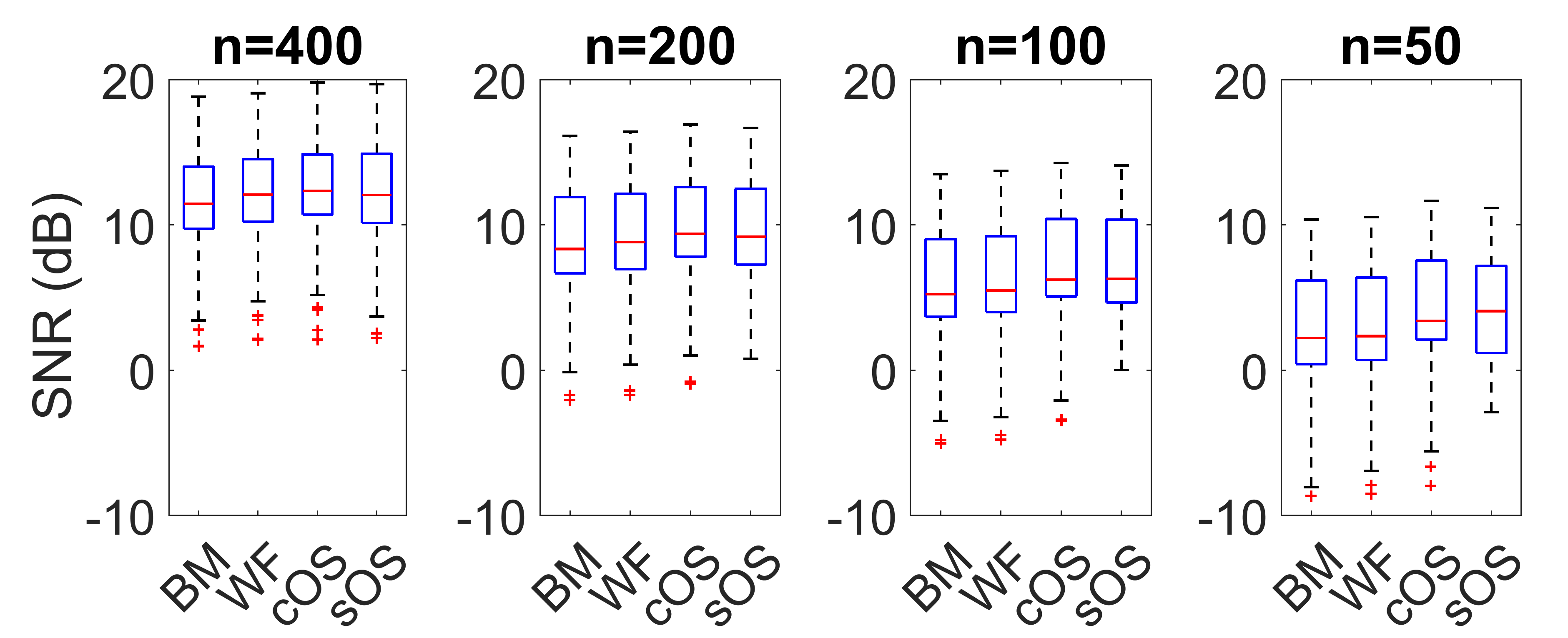}
  \label{fig:nonlinear_SNR}
}\quad
\subfloat[SNR enhancement]{
  \includegraphics[width=9cm]{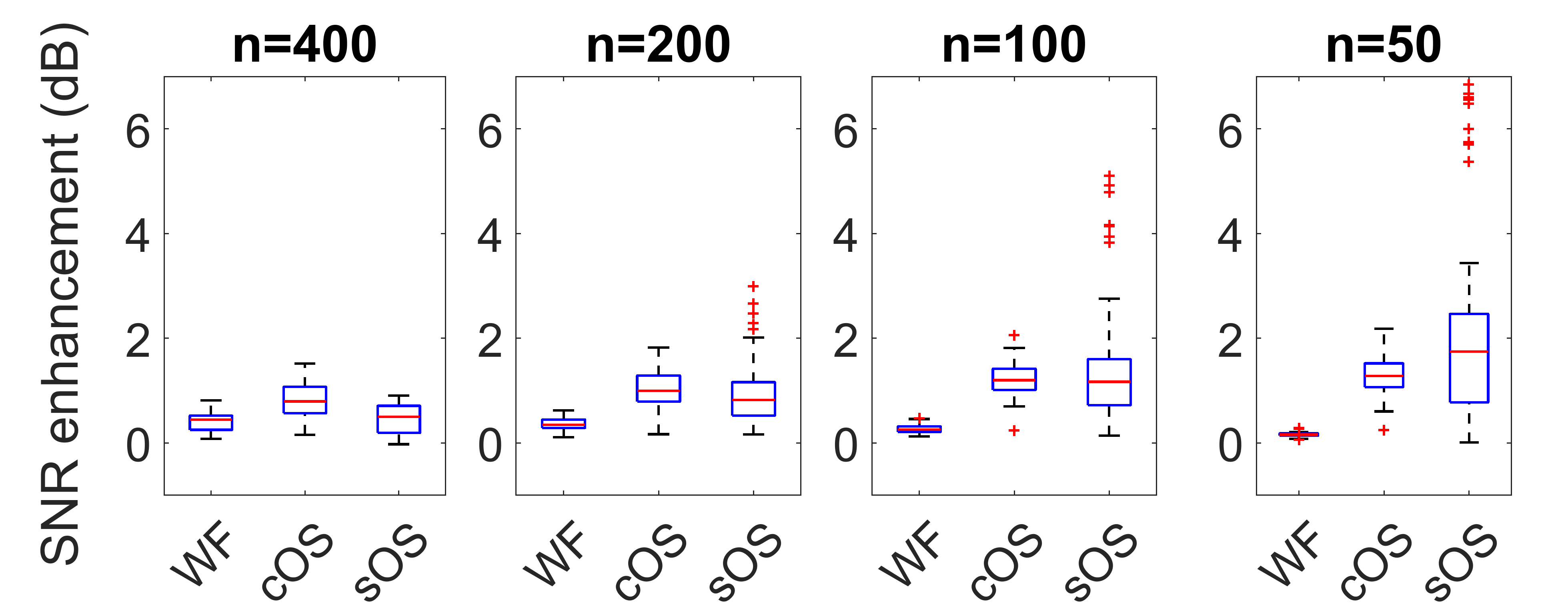}
  \label{fig:nonlinear_SNRen}
}
\caption{\label{fig:nonlinear} The SNR derived from all the 42 recordings. 
(a) the SNRs achieved by different methods, and (b) the corresponding SNR enhancement.
 Labels of the methods are defined the same way as in Table~\ref{tab:result_sim}. 
 }
\end{figure}

Fig.~\ref{fig:nonlinear} summarizes the performance of the denoising methods on the entire dataset for different $n$. By inspection, the variance of SNR across different ears is high and the difference between methods might  appear to be insignificant. 
 Therefore, we define a performance index 
 called ``SNR enhancement'' for comparison purposes; for each of the 42 recording sessions, the SNR enhancement is defined as the difference between the SNR of the signal with or without applying matrix denoising. As shown in Fig.~\ref{fig:nonlinear_SNRen}, the median of SNR enhancement is consistently higher than 0 dB for the three methods, though never greater than 2 dB. However, there are many outliers in the sOS results and the SNR enhancement for some cases is actually higher than 6 dB. The high variance across subjects and the presence of these outliers indicate that the level of SNR enhancement by sOS is not the same for every CEOAE recording. In contrast, the SNR enhancement by the cOS has a narrower range, and its median is higher than the sOS except at $n=50$.

\begin{figure}[ht]
\centering
\subfloat[]{
  \includegraphics[width=6cm]{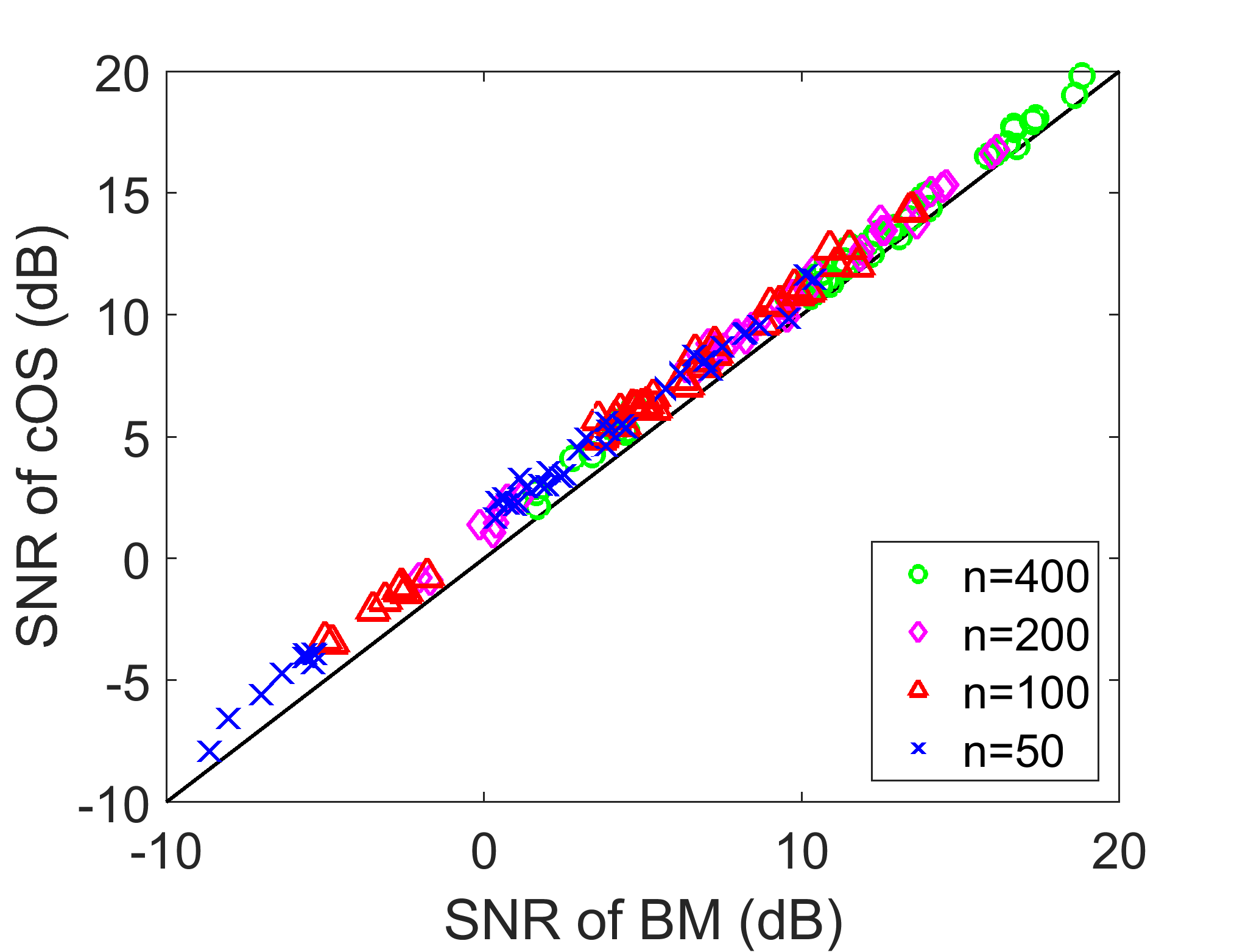}
  \label{fig:snr_vs_wf}
}
\hspace{0cm}
\subfloat[]{
  \includegraphics[width=6cm]{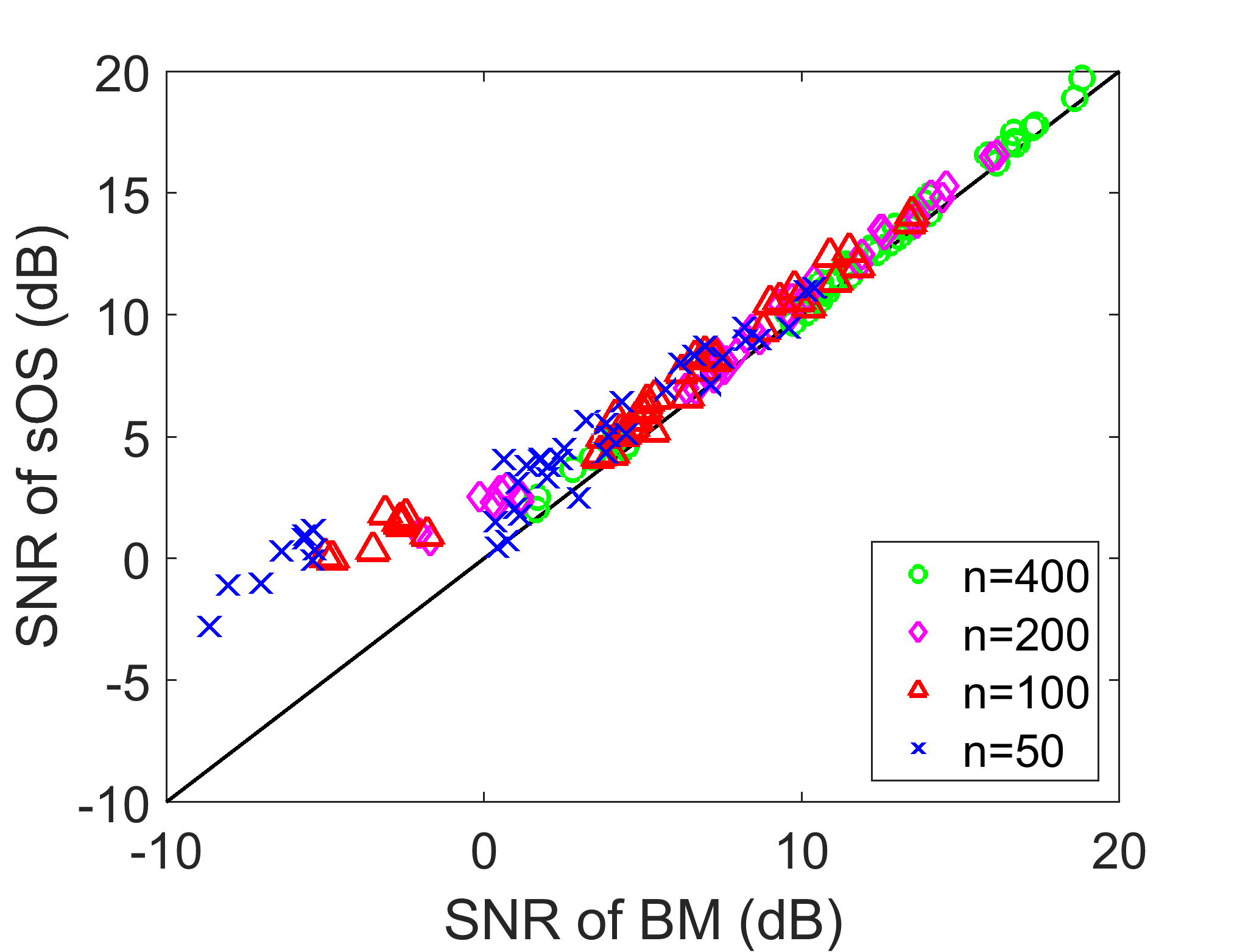}
  \label{fig:snr_vs_os}
}
\caption{\label{fig:snr_vs} The SNR achieved by the baseline method v.s. the matrix denoising methods. (a) cOS. (b) sOS.}
\end{figure}

To investigate under what conditions the sOS would perform well, the SNR achieved by matrix denoising is plotted against the SNR achieved by the BM in Fig.~\ref{fig:snr_vs}. As shown in Fig.~\ref{fig:snr_vs_os}, it becomes clear that the enhancement by the sOS diminishes as the baseline SNR increases; the reduction of noise is more obvious when the baseline SNR is near or below 0 dB. In contrast, the SNR enhancement by the cOS is not affected by the baseline SNR, as shown in Fig.~\ref{fig:snr_vs_wf}. The result indicates that the sOS method can potentially shorten the measurement time in a noisy environment, and this property would be relevant for applying the matrix denoising methods to signals obtained at home or other places outside a sound booth.

\section{Discussion} \label{sec:bigSec4}
In this section, properties of OS on CEOAE signals are first described in signal processing perspective. Afterwards, the estimation of noise is discussed, and the behaviors of sOS on real CEOAEs are inspected. Finally, we discuss the possibility of investigating slow-varying CEOAE dynamics by matrix denoising.

\subsection{Signal processing perspective of optimal shrinkage as a filter}
Although both the amplitude and the frequency fluctuate within 20 ms, the CEOAE can also be regarded as a combination of limited components. If the components of the CEOAE are known, then a ``band-pass filter'' could be applied and the noise could be eliminated. Following this concept, we could determine an adaptive basis of signal by the SVD and the shrinkage on singular values could be understood as a band-pass filter. As proved in \cite{gavish2017}, this method minimizes the loss function between the estimated signal and the desired signal under the Frobenius norm. In the simulated data, the sOS is able to reduce the noise, and the SNR increases over 3 dB at $n\leq 100$, which indicates the recording time could be halved if the sOS is applied. A similar discussion from the signal processing perspective holds for the cOS, while the basis is described by the covariance structure.

The noise model for the cOS and the sOS also deserves a discussion. While the Gaussian white noise is considered in \eqref{eq:YformC} and \eqref{eq:Yform}, it is reasonable to question if noise in CEOAE is white or colored. It is also reasonable to ask if the noise is additive, multiplicative, or even non-linearly related to the signal. To our knowledge, this is a problem without a consensus answer so far. When the noise is colored and additive, there has been some recent work developing OS algorithm under the colored noise model \cite{nadakuditi2014optshrink,leeb2018optimal,leeb2019matrix}. While theoretically the new OS adaptive to colored noise should work better when the noise is colored, however, the improvement is not significant from a very preliminary result in real data. A possibility is that the associated OS algorithm ``overfits'' the colored noise model.\footnote{A private communication with Professor William Leeb.} Solving this color noise issue, and having a further understanding of the noise structure, is an important key toward a better result. 

\subsection{Issues about noise estimation}
Another challenging issue is estimating the noise level $\sigma$, or the rank of signal, in \eqref{eq:YformC} and \eqref{eq:Yform}. There have been several efforts in this direction. See \cite{kritchman2008determining,kritchman2009non,dobriban2019deterministic} for some recent results. While these results work well in other data, their performance is not as expected in this work. One possible reason is that the noise structure in our data is more complicated than the assumed model, so that the estimated $\sigma$ is not ideal. We thus estimate the noise level under the assumption that the recorded CEOAE contains purely noise awhile after the stimulation. While this assumption might sometimes be violated when synchronized spontaneous OAE exists, we found its performance reasonable in real data.

We shall emphasize that to our knowledge, there is no universally accepted solution for noise estimation of CEOAE signals. If we estimate $\sigma$ from the recording outside the ear canal, the noise level could be different from that inside the ear canal. Leaving a few units blank during the measurement for estimating the noise level might be a reasonable strategy if the application scenario allows.

\subsection{Some peculiar behaviors of optimal shrinkage on real CEOAE signals} \label{sec:special_cases}
One of the basic assumptions of OS is that the statistics of noise is stationary in the signal matrix, but this could not be guaranteed in real measurements. For real data, when the SNR attained by the BM is larger than a few dB, the sOS would not further enhance the SNR. The same trend is also observed in the simulation with office or restaurant noise, which is more non-stationary. Nevertheless, the enhancement occurs while the baseline SNR is near or lower than 0 dB. The possible reason could be investigated by comparing the distribution of singular values before and after applying the sOS. Fig.~\ref{fig:singular_real} shows the singular values of file\#2 and file\#10 at $n=100$ before and after the sOS. In this example, the CEOAE signal matrix is with $n=100$ and $p=882$. Due to the nature of OS in Eq.~\eqref{eq:shrinker}, a smaller singular value is reduced even more than a larger one, and the values become all 0 after the 14-th of them. The first singular value of file\#2 is significantly larger than most of others because the SNR is high and the main components can be easily found by the SVD. In contrast, the SNR for file\#10 is bad and its first singular value is not much higher than others. The difference might be a reason that the sOS cannot provide further enhancement while the baseline SNR is higher than 0 dB.

\begin{figure}[htb]
    \centering
    \subfloat[]{
        \includegraphics[width=7cm]{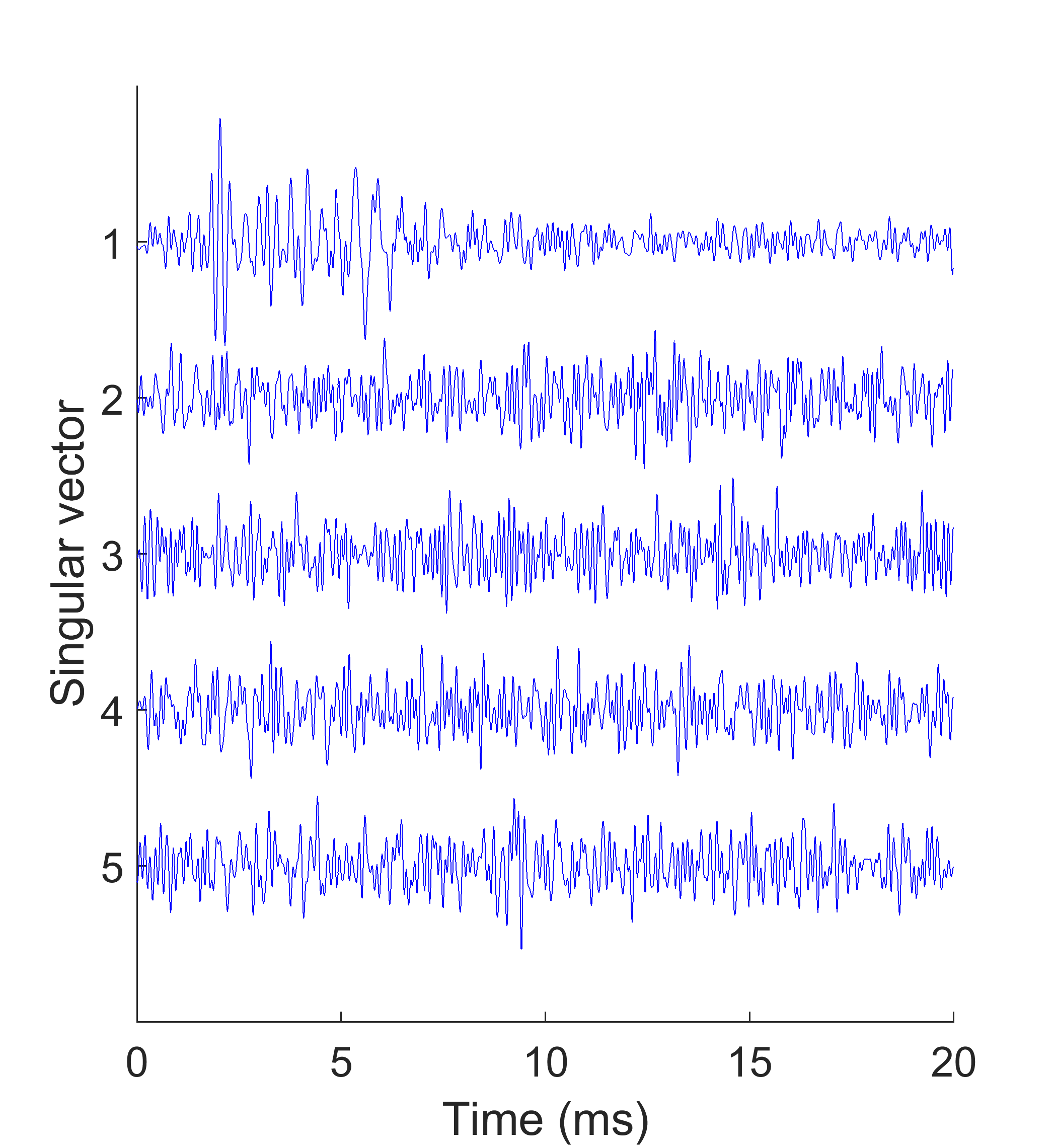}
        \label{fig:singularvector_sim}
    }\quad
    \subfloat[]{
        \includegraphics[width=7cm]{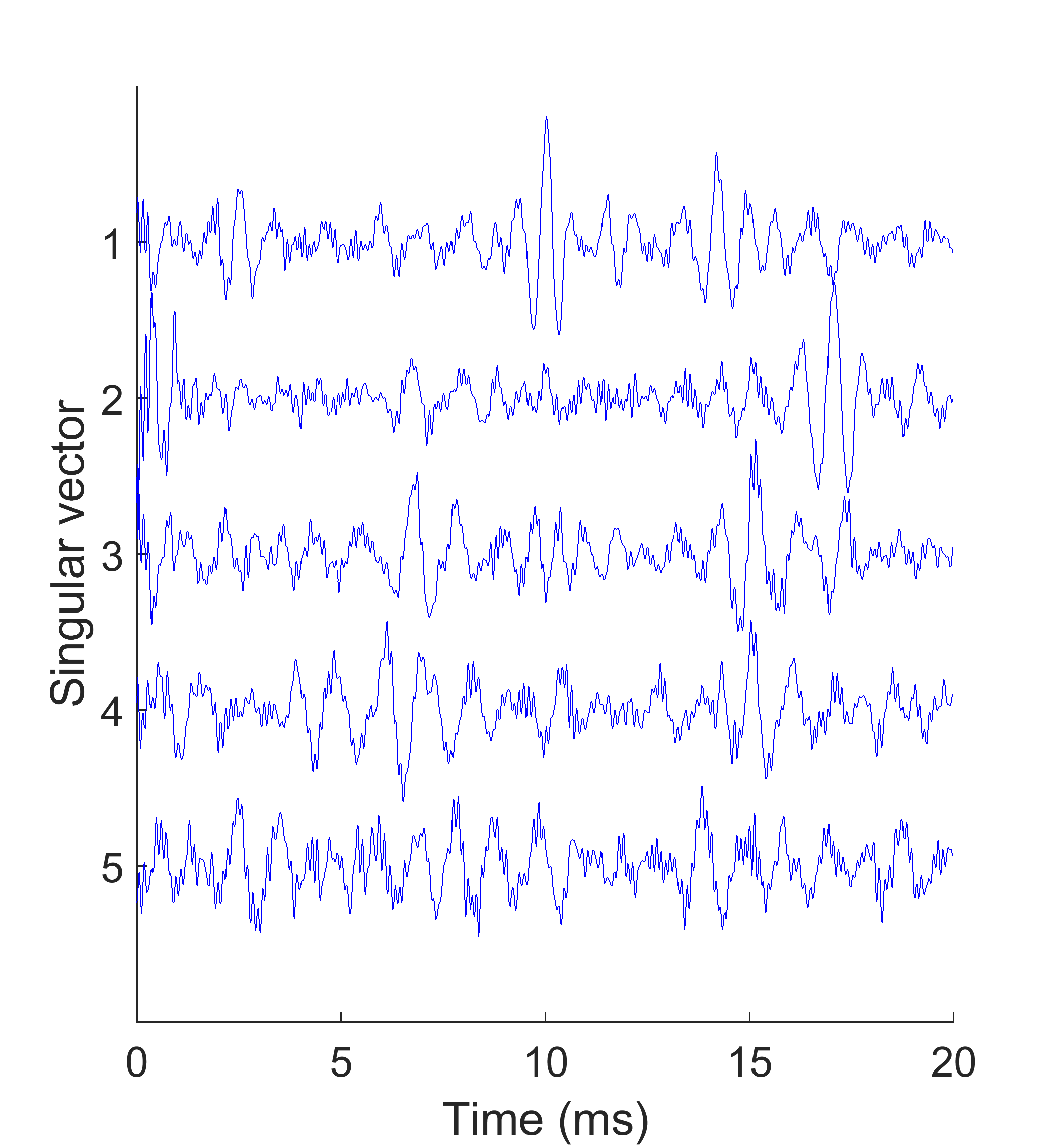}
        \label{fig:singularvector_real_f34}
    }
    \caption{The first five left singular vectors. (a) Simulated CEOAE. (b) The CEOAE of file\#34.}
\end{figure}

We shall emphasize that before the shrinkage, the first singular vector contains not only the desired signal but also the noise. For example, in the simulation shown in Fig.~\ref{fig:singularvector_sim}, the first singular vector contains noise, especially after 15 ms compared to the corresponding ground truth in Fig.~\ref{fig:conceFTA}. The first singular value of file\#2 is big compared with the remaining singular values, so it is not reduced much according to the OS theory. In contrast, the noise in file\#10 is relative large compared with the signal since the first singular value is not dominantly larger than the remaining singular values. Therefore, the denoising effect is stronger for file\#10. This might be why the enhancement by OS diminishes as the SNR increases in real data.

Among the 14 ears, data from one ear show that the amplitude of OS-estimated CEOAE is notably smaller than the desired signal, as shown in Fig.~\ref{fig:ENVreal}e and Fig.~\ref{fig:ENVreal}f. This might be due to a failure of SVD to capture the main components. For this particular example, the first five singular vectors are shown in Fig.~\ref{fig:singularvector_real_f34}. For the simulated CEOAE, the first singular vector is similar to the clean CEOAE in Fig.~\ref{fig:conceFTA} although noise is also included. For a comparison, note that none of the first five singular vectors for file\#34 shown in Fig.~\ref{fig:singularvector_real_f34} is similar to the desired CEOAE shown in Fig.~\ref{fig:ENVreal}. It comes from the fact that the variation of the clean recorded CEOAE signals is large so that we need many singular vectors to describe the signals well. As a result, the rank of the signal matrix is high, and it is more challenging to recover all the associated singular vectors. This fact prevents the sOS from properly eliminating the noise without distorting the desired signal.

\subsection{Investigating the CEOAE dynamics}
Usually the dynamics of CEOAE across units is not considered in CEOAE studies because any subtle change is often masked by the background noise. If the CEOAE is estimated properly, the whole matrix is clarified by matrix denoising methods and the underlying desired signal could be discovered. An example is file\#2 and its waveform is shown in Fig.~\ref{fig:ENVreal}. For this signal, the SNR of a single unit is enhanced from $-6.68\pm 1.91$ dB to $2.47\pm 4.47$ dB by sOS. In such case, the entire matrix instead of only the median is clean and could be analyzed, and the CEOAE dynamics can possibly be observed. The dynamics is a potential feature for the investigation of the auditory system beyond the cochlea, such as to study the contralateral suppression due to activation of the medial olivocochlear reflex pathway (e.g., \cite{Mertes2016}). 

The SVD-based OS and the nature of OAE rings the bell of spatiotemporal analysis \cite{aubry1991spatiotemporal}. Indeed, since the oscillatory frequency reflects the inner ear location, each OAE signal could be viewed as detecting spatial information at different locations inside the inner ear. On the other hand, sequential OAE signals could be viewed as time, which indicates the temporal information of the inner ear. It is thus natural to expect that tools from spatiotemporal analysis society could be applied to further explore inner ear structure. We will explore this possibility in our future work. 

\section{Conclusion} \label{sec:bigSec5}
In this study, CEOAE signals were denoised by Wiener filtering and optimal shrinkage. To verify the methods, 8 subjects were recruited and three CEOAE measurements were conducted for each subject across different days. When applied on these data, the covariance-based OS enhanced the SNR by up to 2 dB compared to the baseline method, and its performance was consistent among different ears. In contrast, the enhancement by the SVD-based OS could be more than 3 dB when the baseline SNR was near or below 0 dB. The results suggested that the cOS can be applied on all CEOAE signals and consistently enhance the SNR, and the sOS can potentially shorten the measurement time and enable CEOAE measurements at home or other places outside a sound booth. Thus, the whole signal matrix was denoised and the SNR for each individual column was enhanced by sOS. This property might allow us to monitor CEOAE dynamics for the purpose of understanding auditory responses beyond the cochlea.

\section*{Acknowledgment}
This research was supported by the Ministry of Science and Technology of Taiwan under grant No. 107-2221-E-007 -093-MY2.

\bibliographystyle{abbrv}
\bibliography{mybib}

\begin{thebibliography}{10}

\bibitem{abdala2009}
C.~Abdala, S.~K. Mishra, and T.~L. Williams.
\newblock Considering distortion product otoacoustic emission fine structure in
  measurements of the medial olivocochlear reflex.
\newblock {\em J. Acoust. Soc. Amer.}, 125(3):1584--1594, 2009.

\bibitem{aubry1991spatiotemporal}
N.~Aubry, R.~Guyonnet, and R.~Lima.
\newblock Spatiotemporal analysis of complex signals\: theory and applications.
\newblock {\em Journal of Statistical Physics}, 64(3-4):683--739, 1991.

\bibitem{biswal2018}
M.~Biswal and S.~K. Mishra.
\newblock Comparison of time-frequency methods for analyzing stimulus frequency
  otoacoustic emissions.
\newblock {\em J. Acoust. Soc. Amer.}, 143(2):626--639, 2018.

\bibitem{Bowling2019}
T.~Bowling, C.~Lemons, and J.~Meaud.
\newblock Reducing tectorial membrane viscoelasticity enhances spontaneous
  otoacoustic emissions and compromises the detection of low level sound.
\newblock {\em Scientific Reports}, 9:7494, 2019.

\bibitem{collet1989}
L.~Collet, M.~Gartner, A.~Moulin, I.~Kauffmann, and F.~Disant.
\newblock Evoked otoacoustic emissions and sensorineural hearing loss.
\newblock {\em Archives of Otolaryngology–Head \& Neck Surgery},
  115(9):1060--1062, 1989.

\bibitem{daubechies2016}
I.~Daubechies, Y.~Wang, and H.~T. Wu.
\newblock Conceft: {C}oncentration of frequency and time via a multitapered
  synchrosqueezed transform.
\newblock {\em Philosophical Transactions of the Royal Society A: Mathematical,
  Physical and Engineering Sciences}, 374:20150193, 2016.

\bibitem{dobriban2019deterministic}
E.~Dobriban and A.~B. Owen.
\newblock Deterministic parallel analysis: an improved method for selecting
  factors and principal components.
\newblock {\em Journal of the Royal Statistical Society: Series B (Statistical
  Methodology)}, 81(1):163--183, 2019.

\bibitem{donoho2018optimal}
D.~L. Donoho, M.~Gavish, and I.~M. Johnstone.
\newblock Optimal shrinkage of eigenvalues in the spiked covariance model.
\newblock {\em Ann. Stat.}, 46(4):1742, 2018.

\bibitem{gavish2017}
M.~Gavish and D.~L. Donoho.
\newblock Optimal shrinkage of singular values.
\newblock {\em {IEEE} Trans. Inf. Theory}, 63(4):2137--2152, 2017.

\bibitem{Goodman2020}
S.~S. Goodman, C.~Lee, J.~J. {Guinan Jr}, and J.~T. Lichtenhan.
\newblock The spatial origins of cochlear amplification assessed by
  stimulus-frequency otoacoustic emissions.
\newblock {\em Biophysical Journal}, 118(5):1183--1195, 2020.

\bibitem{guinan2003}
J.~J. {Guinan Jr}, B.~C. Backus, W.~Lilaonitkul, and V.~Aharonson.
\newblock Medial olivocochlear efferent reflex in humans: otoacoustic emission
  ({OAE}) measurement issues and the advantages of stimulus frequency {OAE}s.
\newblock {\em J. Assoc. Res. Otolaryngol.}, 4(4):521--540, 2003.

\bibitem{hall1999}
A.~J. Hall and M.~E. Lutman.
\newblock Methods for early identification of noise-induced hearing loss.
\newblock {\em Audiology}, 38(5):277--280, 1999.

\bibitem{hotz1993}
M.~A. Hotz, R.~Probst, F.~P. Harris, and R.~Hauser.
\newblock Monitoring the effects of noise exposure using transiently evoked
  otoacoustic emissions.
\newblock {\em Acta Oto-laryngologica}, 113(4):478--482, 1993.

\bibitem{jedrzejczak2012}
W.~W. Jedrzejczak, A.~Bell, P.~H. Skarzynski, K.~Kochanek, and H.~Skarzynski.
\newblock Time–frequency analysis of linear and nonlinear otoacoustic
  emissions and removal of a short-latency stimulus artifact.
\newblock {\em J. Acoust. Soc. Amer.}, 131(3):2200--2208, 2012.

\bibitem{jedrzejczak2008}
W.~W. Jedrzejczak, K.~J. Blinowska, K.~Kochanek, and H.~Skarzynski.
\newblock Synchronized spontaneous otoacoustic emissions analyzed in a
  time-frequency domain.
\newblock {\em J. Acoust. Soc. Amer.}, 124(6):3720--3729, 2008.

\bibitem{johnstone2006high}
I.~M. Johnstone.
\newblock High dimensional statistical inference and random matrices.
\newblock In {\em Proceedings oh the International Congress of Mathematicians:
  Madrid, August 22-30, 2006: invited lectures}, pages 307--333, 2006.

\bibitem{kemp1978}
D.~T. Kemp.
\newblock Stimulated acoustic emissions from within the human auditory system.
\newblock {\em J. Acoust. Soc. Amer.}, 64(5):1386--1391, 1978.

\bibitem{kemp1986_2}
D.~T. Kemp.
\newblock Otoacoustic emissions, travelling waves and cochlear mechanisms.
\newblock {\em Hearing Research}, 22(1-3):95--104, 1986.

\bibitem{kemp1986}
D.~T. Kemp, P.~Bray, L.~Alexander, and A.~M. Brown.
\newblock Acoustic emission cochleography--practical aspects.
\newblock {\em Scandinavian Audiology. Supplementum}, 25:71--95, 1986.

\bibitem{kritchman2008determining}
S.~Kritchman and B.~Nadler.
\newblock Determining the number of components in a factor model from limited
  noisy data.
\newblock {\em Chemometrics and Intelligent Laboratory Systems}, 94(1):19--32,
  2008.

\bibitem{kritchman2009non}
S.~Kritchman and B.~Nadler.
\newblock Non-parametric detection of the number of signals: Hypothesis testing
  and random matrix theory.
\newblock {\em IEEE Trans. Signal Process.}, 57(10):3930--3941, 2009.

\bibitem{leeb2019matrix}
W.~Leeb.
\newblock Matrix denoising for weighted loss functions and heterogeneous
  signals.
\newblock {\em arXiv preprint arXiv:1902.09474}, 2019.

\bibitem{leeb2018optimal}
W.~Leeb and E.~Romanov.
\newblock Optimal spectral shrinkage and {PCA} with heteroscedastic noise.
\newblock {\em arXiv preprint arXiv:1811.02201}, 2018.

\bibitem{liu2016}
Y.~W. Liu and T.~C. Liu.
\newblock Quasilinear reflection as a possible mechanism for suppressor-induced
  otoacoustic emission.
\newblock {\em J. Acoust. Soc. Amer.}, 140(6):4193--4203, 2016.

\bibitem{liu2009}
Y.~W. Liu and S.~T. Neely.
\newblock Outer hair cell electromechanical properties in a nonlinear
  piezoelectric model.
\newblock {\em J. Acoust. Soc. Amer.}, 126(2):751--761, 2009.

\bibitem{liu2010}
Y.~W. Liu and S.~T. Neely.
\newblock Distortion product emissions from a cochlear model with nonlinear
  mechanoelectrical transduction in outer hair cells.
\newblock {\em J. Acoust. Soc. Amer.}, 127(4):2420--2432, 2010.

\bibitem{MoH2014}
Y.~W. Liu, L.~M. Yu, and P.~J. Wu.
\newblock Close-loop simulation of the medial olivocochlear anti-masking
  effects.
\newblock In K.~D. Karavitaki and D.~P. Corey, editors, {\em Mechanics of
  Hearing: Protein to Perception}, page 090028. American Institute of Physics,
  2015.

\bibitem{marchesi2013}
S.~Marchesi, G.~Togonola, and A.~Paglialonga.
\newblock A bispectral approach to analyze nonlinear cochlear active mechanisms
  in transient evoked otoacoustic emissions.
\newblock {\em {IEEE} Trans. Biomed. Circuits Syst.}, 7(4):401--403, 2013.

\bibitem{Mertes2016}
I.~B. Mertes and S.~S. Goodman.
\newblock Within- and across-subject variability of repeated measurements of
  medial olivocochlear-induced changes in transient-evoked otoacoustic
  emissions.
\newblock {\em Ear and Hearing}, 37(2):e72--e84, 2016.

\bibitem{moleti2005}
A.~Moleti, R.~Sisto, G.~Tognola, M.~Parazzini, P.~Ravazzani, and F.~Grandori.
\newblock Otoacoustic emission latency, cochlear tuning, and hearing
  functionality in neonates.
\newblock {\em J. Acoust. Soc. Amer.}, 118(3):1576--1584, 2005.

\bibitem{mountain1994}
D.~C. Mountain and A.~E. Hubbard.
\newblock A piezoelectric model of outer hair cell function.
\newblock {\em J. Acoust. Soc. Amer.}, 95(1):350--354, 1994.

\bibitem{nadakuditi2014optshrink}
R.~R. Nadakuditi.
\newblock Optshrink: An algorithm for improved low-rank signal matrix denoising
  by optimal, data-driven singular value shrinkage.
\newblock {\em IEEE Trans. Inf. Theory}, 60(5):3002--3018, 2014.

\bibitem{neely1985}
S.~T. Neely.
\newblock Mathematical modeling of cochlear mechanics.
\newblock {\em J. Acoust. Soc. Amer.}, 78(1):345--352, 1985.

\bibitem{neely1993}
S.~T. Neely.
\newblock A model of cochlear mechanics with outer hair cell motility.
\newblock {\em J. Acoust. Soc. Amer.}, 94(1):137--146, 1993.

\bibitem{notaro2007}
G.~Notaro, A.~M. Al-Maamury, A.~Moleti, and R.~Sisto.
\newblock Wavelet and matching pursuit estimates of the transient-evoked
  otoacoustic emission latency.
\newblock {\em J. Acoust. Soc. Amer.}, 122(6):3576--3585, 2007.

\bibitem{onorati2013}
R.~Onorati, P.~Sampson, and P.~Guttorp.
\newblock A spatio-temporal model based on the {SVD} to analyze daily average
  temperature across the {S}icily region.
\newblock {\em Journal of Environmental Statistics}, 5, 2013.

\bibitem{prieve1995}
B.~A. Prieve and S.~R. Falter.
\newblock {COAE}s and {SSOAE}s in adults with increased age.
\newblock {\em Ear and Hearing}, 16(5):521--528, 1995.

\bibitem{ravazzani1993}
P.~Ravazzani and F.~Grandori.
\newblock Evoked otoacoustic emissions: nonlinearities and response
  interpretation.
\newblock {\em {IEEE} Trans. Biomed. Eng.}, 40(5):500--504, 1993.

\bibitem{ravazzani1998}
P.~Ravazzani, G.~Tognola, F.~Grandori, and J.~Ruohonen.
\newblock Two-dimensional filter to facilitate detection of transient-evoked
  otoacoustic emissions.
\newblock {\em {IEEE} Trans. Biomed. Eng.}, 45(9):1089--1096, 1998.

\bibitem{ravazzani2003}
P.~Ravazzani, G.~Tognola, M.~Parazzini, and F.~Grandori.
\newblock Principal component analysis as a method to facilitate fast detection
  of transient-evoked otoacoustic emissions.
\newblock {\em {IEEE} Trans. Biomed. Eng.}, 50(2):249--252, 2003.

\bibitem{shera2001}
C.~A. Shera.
\newblock Intensity\-invariance of fine time structure in basilar\-membrane
  click responses\: Implications for cochlear mechanics.
\newblock {\em J. Acoust. Soc. Amer.}, 110(1):332--348, 2001.

\bibitem{singer2013}
A.~Singer and H.~T. Wu.
\newblock Two-dimensional tomography from noisy projections taken at unknown
  random directions.
\newblock {\em SIAM Journal on Imaging Sciences}, 6(1):136--175, 2013.

\bibitem{sisto2007}
R.~Sisto and A.~Moleti.
\newblock Transient evoked otoacoustic emission latency and cochlear tuning at
  different stimulus levels.
\newblock {\em J. Acoust. Soc. Amer.}, 122(4):2183--2190, 2007.

\bibitem{sisto2019}
R.~Sisto, C.~A. Shera, A.~Altoè, and A.~Moleti.
\newblock Constraints imposed by zero\-crossing invariance on cochlear models
  with two mechanical degrees of freedom.
\newblock {\em J. Acoust. Soc. Amer.}, 146(3):1685--1695, 2019.

\bibitem{talmadge1998}
C.~L. Talmadge, A.~Tubis, G.~R. Long, and P.~Piskorski.
\newblock Modeling otoacoustic emission and hearing threshold fine structures.
\newblock {\em J. Acoust. Soc. Amer.}, 104(3):1517--1543, 1998.

\bibitem{tognola1997}
G.~Tognola, F.~Grandori, and R.~Ravazzani.
\newblock Time-frequency distributions of click-evoked otoacoustic emissions.
\newblock {\em Hearing Research}, 106(1-2):112--122, 1997.

\bibitem{tognola2005}
G.~Tognola, M.~Parazzini, P.~{De Jager}, P.~Brienesse, P.~Ravazzani, and
  F.~Grandori.
\newblock Cochlear maturation and otoacoustic emissions in preterm infants: a
  time–frequency approach.
\newblock {\em Hearing Research}, 199(1-2):71--80, 2005.

\bibitem{Vencovsky2020}
V.~Vencovsk\'{y}, A.~Vete\v{s}n\'{i}k, and A.~W. Gummer.
\newblock Nonlinear reflection as a cause of the short-latency component in
  stimulus-frequency otoacoustic emissions simulated by the methods of
  compression and suppression.
\newblock {\em J. Acoust. Soc. Amer.}, 147:3992--4008, 2020.

\bibitem{Verhulst2012}
S.~Verhulst, T.~Dau, and C.~A. Shera.
\newblock Nonlinear time-domain cochlear model for transient stimulation and
  human otoacoustic emission.
\newblock {\em J. Acoust. Soc. Amer.}, 132:3842--3848, 2012.

\bibitem{wiener1950}
N.~Wiener.
\newblock {\em Extrapolation, interpolation, and smoothing of stationary time
  series: with engineering applications}.
\newblock MIT press, 1950.

\bibitem{withnell2005}
R.~H. Withnell and S.~McKinley.
\newblock Delay dependence for the origin of the nonlinear derived transient
  evoked otoacoustic emission.
\newblock {\em J. Acoust. Soc. Amer.}, 117(1):281--291, 2005.

\bibitem{wu2018}
H.~T. Wu and Y.~W. Liu.
\newblock Analyzing transient-evoked otoacoustic emissions by concentration of
  frequency and time.
\newblock {\em J. Acoust. Soc. Amer.}, 144(1):448--466, 2018.

\end{thebibliography}

\end{document}